\documentclass[aps,twocolumn, preprintnumbers]{revtex4}

\usepackage{etoolbox}
\makeatletter
\patchcmd{\frontmatter@RRAP@format}{(}{}{}{}
\patchcmd{\frontmatter@RRAP@format}{)}{}{}{}
\renewcommand\Dated@name{}
\makeatother

\usepackage{graphicx}
\usepackage{subfigure}
\usepackage{multirow}
\usepackage{fancyhdr}
\usepackage{longtable}
\usepackage{parskip}
\usepackage[T1]{fontenc}
\usepackage{dcolumn}  

\usepackage{bm}       
\usepackage{amsfonts}
\usepackage{amsmath}  
\usepackage{amssymb}

\newcommand{\pwisein}{\left\{ \begin{array}{ll}}
\newcommand{\pwiseout}{\end{array}\right.}

\begin{document}

\title{On Regular Negative Mass Black Holes Under Unitary Time and Proper Antichronous Transformations}

\author{Manuel Urueña Palomo}

\affiliation {\it Technical University of Munich (Germany), ge58vib@mytum.de, muruep00@gmail.com}

\date{September, 2024}

\begin{abstract}  

A non-singular black hole solution is briefly presented which violates energy conditions only at its interior by postulating a consistent shift to negative energies and gravitationally repulsive negative masses at the event horizon. This shift is the unitary parity-time $PT$ transformation of relativistic quantum mechanics and the proper antichronous transformation of the full Lorentz group. The transformation at the event horizon respects Einstein's equivalence principle, and the considered negative mass interaction does not result in the runaway motion paradox or vacuum instability. The correspondence of these regular black holes with observed ones is studied by proposing another mechanism of black hole growth independent of accretion and merging due to interior increasing entropy, which attempts to solve the unexplained size and formation of high-redshift supermassive black holes, the intermediate mass gap, and the information paradox.

\end{abstract}

\maketitle 

{\bfseries Keywords:}
    black holes; singularities; negative mass; energy conditions; unitary time transformation; antichronous transformations; supermassive black holes

\section*{ \bfseries 1. Introduction}

\hspace{5mm} Current description of black holes (BHs) relies on singularities enclosed by event horizons (EHs), which are the past causal boundary of future null infinity in asymptotically flat spacetime, i.e., the surface around BHs under which no luminal or subluminal entities or signals can escape outside due to gravity. This follows from the standard Schwarzschild solution in General Relativity (GR) and Penrose singularity theorems applied to the processes of stellar gravitational collapse. 

\hspace{5mm} Evidence for the existence of BH’s EHs is found in X-ray binary systems with massive non-bright partners, star orbits around non-bright sources, X-ray emissions from accretion disks and gravitational wave signals from mergers, among others.

\hspace{5mm} The observed mass of supermassive black holes (SMBHs) at redshifts of $z\sim10$ cannot be currently explained by the known BH growth processes: accretion is limited by the Eddington mass limit and merging by the final parsec problem. Primordial BHs and processes of direct collapse have been proposed in order to explain the formation and size of these SMBHs, although there is no consensus on their origin.

\hspace{5mm} The inner description of BHs has not been observed or proven to match GR predictions. It relies on the assumption that curvature can be arbitrarily small at EHs and beyond, so that GR as we know it should still be valid and give correct predictions for the interior because it does them in its classical domain. In principle, there is nothing locally distinguished at EHs. Alternative inner descriptions can be hypothesized by defying the assumptions of the interior solution.

\subsection*{ 1.1. Is the gravitational singularity a problem?}
 
\hspace{5mm} Gravitational singularities are problematic in GR because they imply geodesic incompleteness. It is an open question whether gravitational singularities exist as predicted by GR, or if a more complete description of gravity, such as a quantum theory of gravity, solves them. Quantum gravity effects would be non-negligible when quantum fluctuations become too large at very high mass densities and spacetime curvature reaches values out of the domain to which GR can effectively be applied. This has led the majority of physicists to assume that the current description of BH interiors in GR is accurate up to very high curvature regions near the singularity. 

\hspace{5mm} Thus, these singularities are not thought to be problematic, but this should not stop the search for alternative interior descriptions, as they are part of the universe in which experimental physics could be performed (although results could not be communicated to the exterior) and physics attempts to fundamentally understand natural phenomena. In particular, modifications to GR might solve gravitational singularities without invoking the still elusive quantum gravity theory.

\hspace{5mm} Regular BHs are solutions similar to the standard Schwarzschild solution that are singularity free. Examples of regularization of the standard Schwarzschild solution involve including magnetic matter fields in the energy-momentum tensor so that the singularity is replaced by a De Sitter space core with false vacuums, changing the mass parameter to a radial mass function, coupling the field equations to non–linear electromagnetic sources, limiting curvatures, using different or extended metric theories of gravity, etc. Although there are many mathematical ways to regularize a BH interior solution, there are no theoretical motivations or physical reasons why the cited options actually take place inside observed BHs.

\subsection*{1.2. Negative mass black holes}

\hspace{5mm} Eternal negative mass BHs have been considered unphysical because they would exhibit naked singularities. Initial negative mass configurations would interact with nearby positive masses, leading to the runaway motion. Extensive research has been conducted on negative mass BHs, but only considering either an entire negative mass universe \cite{Bonnor1989}, alternative negative mass interactions to prevent the runaway motion paradox between unlike masses \cite{Petit2015}, a change of coordinates together with mass inversion to cancel the repulsive negative mass interactions \cite{Bondarenko2019}, or the collapse of a violating energy condition initial matter configuration leading to a negative mass BH \cite{Mann1997}, and \cite{BELLETTE2013} which results in the runaway paradox.

\hspace{5mm} A few lines of research have explored transitioning negative repulsive masses, but only either at the $-r$ region of the Kerr BH \cite{Villata2015}, or at the EH of Schwarzschild BHs by a parity and time transformations \cite{Petit2015} but considering a bimetric model and negative masses attractive under an unjustified choice of negative mass interactions (like masses attract, unlike masses repel) which contradicts Newtonian mechanics and GR predictions. Although $PT$ symmetry transformations leading to negative masses in BHs have been studied in \cite{Bondarenko2019}, negative masses are assumed to maintain their original dynamics, interacting attractively with positive inertial mass, so that a mass inversion together with a $PT$ transformation is used to cancel the repulsive interaction. Time running backwards inside BHs has been considered in \cite{Bousso2015}.

\section*{\bfseries 2. Motivation}
\subsection*{2.1 Penrose singularity theorems}\label{subsec:Hawking-Penrose singularity theorems}

 \hspace{5mm} Penrose singularity theorems \cite{HawkingPenrose1970} state that, if energy conditions are satisfied in closed trapped surfaces, such as the weak energy condition (light rays are always focused together by gravity from non-negative energy of matter), gravitational singularities are unavoidable and spacetime is causally-geodesically incomplete, given a zero cosmological constant. Although energy conditions are reasonable and satisfied in the macroscopic universe, there are quantum effects that seem to violate them, such as quantum fluctuations and the Casimir effect. Considering a macroscopic spacetime with negative energy density (negative masses or exotic matter) would imply these violations, and may avoid the formation of singularities. The dominant energy condition, violated by a non-negative local energy density, is taken as an assumption in the positive energy theorem derivation \cite{Witten1981}, and a non-negative energy of the self coupled gravitational field is possible if spacetime is not asymptotically flat \cite{Deser1968}.

\subsection*{2.2 Negative mass interaction}\label{subsec:negative_mass}

\hspace{5mm} Negative mass has been theoretically studied since Newton’s law of gravity due to its similarity with Coulomb’s law, that applies to different sign charges. Although negative masses have always been regarded as unphysical due to their lack of observational evidence and theoretical problems, one could hypothesize their existence in unobserved regions of spacetime to explain observed phenomena while avoiding the runaway paradox. 

\hspace{5mm} Newton’s law of gravity for two negative masses results in the same forces as the positive masses case, but considering Newton’s second law of motion results in opposite accelerations, and thus, antigravity. This follows from the equivalence principle (EP), which states that gravitational masses are equal to inertial masses. Problems arise when considering the interaction between two masses of different sign, in which the negative mass chases the positive mass, accelerating both presumably up to the speed of light \cite{Luttinger1951}, which has been regarded as the paradox of the runaway motion \cite{Bonnor1989}, \cite{Bondi1957}. Although for any value of mass, both energy and momentum are conserved due to considering negative energy and momentum, this interaction gives rise to paradoxes like infinite energy sources and perpetual motion machines.

\hspace{5mm} In GR, choosing the active gravitational mass in the standard Schwarzschild solution negative results in the approximation of the Newtonian case of repulsion for test particles \cite{Bondi1957}. As a constant of integration, the active gravitational mass can take any arbitrary sign, and GR does not prohibit the existence of negative masses \cite{Luttinger1951}. These negative masses (or exotic matter) have been used to support solutions such as traversable wormholes and warp drives, which also give rise to paradoxes like time travel machines, faster than light travel, or other causality-breaking phenomena. Moreover, if some masses repel others while attracting other ones, this could be used to locally measure the gravitational field, in violation of the EP. Bonnor showed that these problems, such as violations of causality, or equivalently, the violation of the second law of thermodynamics, occur “\textit{only if both positive and negative masses are present}” \cite{Bonnor1989}.

\hspace{5mm} Another argument against negative mass follows that if negative mass coexist with positive mass and the EP holds, they would trigger a vacuum decay, because negative energies are forced to exist and pairs of negative and positive energies would emerge from a net zero energy space in infinite numbers, making the vacuum unstable. What this shows is only that the interaction between positive and negative masses is problematic.

\hspace{5mm} Although there are proposals in which the EP only holds up to a sign, such as in bimetric theories of gravity, the arguments presented before (especially the self-accelerating runaway motion) are usually used to support the idea that negative masses should not follow the Newtonian interaction for gravitational masses being equal to inertial masses, but an analogous interaction to the electrostatic force: like masses attract and unlike masses repel. This interaction follows from the fact that particles with equal charge exchanging a particle of odd spin experience a force pointing away from each other, while particles with equal charge exchanging a particle of even spin experience a force pointing towards each other. The second case would be the case of gravity being mediated by a spin two boson (graviton) as the propagator, with the charge being the energy-momentum of the particle, and equivalently, its mass. This interaction, which is the same as considering inertial masses always positive and allowing passive and active gravitational masses to be negative in Newtonian gravity, and thus, giving up the EP, matches the interaction of a spin two field from Quantum Field Theory, although there is no experimental evidence for gravity to be an interaction mediated by a graviton or a quantum field.

\hspace{5mm} As shown before, all theoretical problems of negative masses only take place regarding the interaction between positive and negative masses, for the case in which the EP holds and negative masses repel each other. This was early realized by Bonnor \cite{Bonnor1989}, who conceived a whole negative mass universe for consistency of his study of negative mass. As Bonnor  pointed out, his universe had “\textit{no practical relation to the one we live in}”, but there is a consistent way to propose the existence of these negative repulsive masses within our universe respecting the EP and with no theoretical inconsistencies: if only negative masses and energies are allowed to exist within the EH of BHs, no interaction could possibly take place with exterior positive masses, and the properties of BHs would naturally impede the problematic runaway motion interaction. Their causal isolation would prevent any potential annihilation resulting from opposite energy signs, and the lack of observational evidence would be explained. Positive and negative energy states can coexist in two separated spacetimes by an EH with increasing energy potential from both sides of a common ground state of lowest energy so that causality and absolute values of energy are conserved. As explained before, gravitational singularities would be prevented by violating energy conditions. Thus, the described model is the only consistent possibility for the existence of negative masses in our universe following the EP.

\hspace{5mm} An interesting feature of negative inertial masses is that acceleration is opposite to a force applied to them: pushing a negative mass makes it accelerate towards the force, while pulling it makes it accelerate away. This strange phenomenon is simply the classical time reversal of the process of applying a force to a positive mass (since force is an even variable that does not change sign under classical time reversal), which already hints a relationship between time transformations and negative masses, which will be presented in the following sections.

\subsection*{2.3. Negative mass transformations in relativistic quantum mechanics} \label{subsec:negative_mass_transformations_in_relativistic_qunatum_mechanics}

\hspace{5mm} In order to propose that negative masses exist inside BHs, a mechanism switching from observed positive mass stars prior to their collapse to negative masses once crossed the EH must postulated. As explained in the following section, this must be a unitary time transformation.

\hspace{5mm}A simple example would be considering the covariant equation of motion or Lorentz force of a charged particle in an electromagnetic field

\begin{equation}
	\frac{d^2 x^\mu}{d\tau^2}= \frac{q}{m}F^\mu _\nu \frac{dx^\nu}{d\tau}
\end{equation}

\hspace{5mm} where $x$ is the position and $F^\mu _\nu$ the electromagnetic field tensor, in which changing the sign of proper time $\tau$ has the same effect as changing the sign of charge $q$, or equivalently, of mass $m$.

\hspace{5mm} Another approach relies on the explanation for antiparticles in the Dirac and Klein-Gordon equation solutions, which is achieved through the Feynman-Stüeckelberg interpretation considering the negative energy solutions of the Dirac equation with phase factor $-i(Et-xp)=i(xp - Et)$ as particles moving backwards in time, so that a change from negative time to positive time $-t \rightarrow t$ is equivalent to a change from negative energies to positive energies $-E \rightarrow E$, and time is also considered to shift momentum $p \rightarrow -p$ for consistency of the phase factor, resulting in positive energy particles $i(Et - xp)$ behaving as having opposite charge.

\hspace{5mm} Thus, negative energy solutions with negative time (particles moving backwards in time) can be interpreted as positive energy solutions with positive time (particles moving forward in time), proving that time transformations change the sign of a particle’s energy in relativistic quantum mechanics. Mathematically, however, there is no reason to reject the negative energy and mass solutions, which could have a real physical correspondence. It is worth noting that this negative mass solution is not antimatter, since antiparticles have positive energy and mass, consistent with the positive energy outcome observed during the annihilation of a matter and antimatter at the positive time region of the universe. 

\hspace{5mm} Time can easily be reversed in classical physics and non-relativistic quantum mechanics (the common notion of a film played backwards), and it implies $t\rightarrow-t$ and $p\rightarrow-p$, but in relativity, it implies moving “backwards” in a dimension which is timelike, for which its effects have never been observed, and thus, it cannot be assumed to be antiunitary. Considering $T$ unitary would imply changing the notion of a time transformation and time symmetry, i.e., it would only switch to negative time $t\rightarrow-t$ and negative energies $E\rightarrow-E$, and would not reverse momentum, so that $TpT^{-1}= p$, and the Feynman-Stüeckelberg interpretation is reduced to a simple change in the sense of $i$ (an antiunitary $PT$ transformation where momentum and energy do not shift signs).

\hspace{5mm} The possibility of $T$ being unitary has been extensively studied in \cite{Debergh2022}. Considering $T$ a linear and unitary operator, it not only changes the sign of energy as also explained for quantum field theory in \cite{Weinberg1995}: “\textit{If we supposed that T is linear and unitary $[\cdots]$ for any state $\psi$ of energy $E$ there is another state $T^{-1} \ \psi$ of energy $-E$. To avoid this, we are forced here to conclude that $T$ is antilinear and antiunitary}”, but also, for a unitary $PT$, it switches the sign of the mass of a particle at rest. This has been shown in further development of unitary $PT$ transformations in \cite{Debergh2018}, in relation to the chirality operator or gamma-five matrix $\gamma^5$, which is associated to reversing the sign of the mass term in the Standard Model Lagrangian \cite{Ashok1994}, \cite{Tiomno1955}. The same matrix $\gamma^5$ represents mass symmetry and is naturally connected to a unitary $PT$ transformation. The same reasoning has been followed in \cite{Marsch2021}. Both energy and mass must be negative in order to have positive density of probabilities in the Klein-Gordon equation by the ratio $E/m$ (the current interpretation of quantum field theory forbids negative energy states since it arbitrarily sets the zero-point field of the quantum vacuum as a ground state of lowest possible energy), and all physical predictions of quantum electrodynamics are independent of the sign of the mass in the Dirac equation \cite{Tiomno1955}.

\hspace{5mm} For the well known $CP$ symmetry violation, the preservation of $CPT$ symmetry requires the violation of $T$ symmetry. In the proposed method, time symmetry breaking implies changing the gravitational interaction from attractive to repulsive by changing the sign of energy, and consequently the sign of mass (or mass-energy) conjugation $M$. Thus, a new broken symmetry $M$ is defined and the resulting $CPTM$ symmetry is not fully invariant. This symmetry group has been considered invariant in \cite{Bondarenko2019} and in \cite{Marsch2021}, by considering $T$ antiunitary.

\hspace{5mm} The Dirac equation applies only to fermions. For consistency of the proposal, all particles must shift their energy by a time transformation. In the case of bosons, obtaining negative energies requires considering either a negative frequency or time ($[s^{-1}]$) or a negative momentum (negative wavelength $\lambda$ $[m]$) while preserving the sign of constants in the equations $E = hf$ and $E = pc = \frac{hc}{\lambda}$ (so that $c= f \lambda$). Negative frequencies can be explained by time transformations, and negative momenta by parity transformations so that $c = (-f)(-\lambda) = f \lambda$. This supports the idea that a unitary time reversal must be accompanied by a parity transformation, so that their combination changes momentum and time $(p_0, t_0) \rightarrow (-p_0, -t_0)$ and energy $E \rightarrow -E$.

\hspace{5mm} The unitary time transformation together with a parity transformation provide an explanation for the existence of solely negative masses and energies within BHs if the transformations take place at EHs.

\subsection*{2.4. Negative mass transformations in Special Relativity}\label{subsec:negative_mass_transformations_in_special_relativity}

\hspace{5mm} The time transformation resulting in negative masses explained in the previous section within the context of relativistic quantum mechanics agrees with the proper antichronous (or non-orthochronous) transformation of the full Lorentz group within Special Relativity (SR).

\hspace{5mm} The proper antichronous transformations $\Lambda_{+}^{\downarrow}$ of the full Lorentz group

\begin{equation}
    \begin{Bmatrix} 
    \{\Lambda_{+}^{\uparrow}\}: \Lambda_0^{0'} \geq +1; \ det \Lambda = +1 \\
    \{\Lambda_{+}^{\downarrow}\}: \Lambda_0^{0'} \leq -1; \ det \Lambda = +1 \\
    \{\Lambda_{-}^{\uparrow}\}: \Lambda_0^{0'} \geq +1; \ det \Lambda = -1 \\
    \{\Lambda_{-}^{\downarrow}\}: \Lambda_0^{0'} \leq -1; \ det \Lambda = -1 
    \end{Bmatrix}
\end{equation}

\hspace{5mm} \noindent are arbitrarily thought to be non-physical, and SR is based on the proper orthochronous transformations or restricted Lorentz group $L _{+}^{\uparrow} \equiv \Lambda _{+}^{\uparrow}$ in $SO(1,3)^{\uparrow}$, dealing only with positive energies and positive times. In the full Lorentz group (2), $\pm$ refers to the sign of the determinant of $\Lambda$ and $\uparrow \downarrow$to the sign of the first element of $\Lambda$.

\hspace{5mm} Proper antichronous transformations can be postulated if SR is based in the whole proper subgroup of proper orthochronous and antichronous Lorentz transformations $L_{+} = L_{+}^{\uparrow} \ \cup \ L_{+}^{\downarrow} \equiv \Lambda_{+}^{\uparrow} \ \cup \ \Lambda_{+}^{\downarrow}$ which forms an invariant subgroup for which ordinary relativistic laws are covariant. As a result of considering the proposed transformations in the previous section, the full Lorentz group should be considered from $(\Lambda_0^{0'})^2 \geq 1$ without the condition of $\Lambda_0^{0'} \geq 1$, with linear and unitary operators describing the improper time reversal $T = (diag(-1, 1, 1, 1))$ and improper parity reversal $P = (diag(1, -1, -1, -1))$, with the four component coordinate position $C = (ct, x, y, z)$, and thus, a proper antichronous transformation $PT = (-diag(1, 1, 1, 1))$ which is equivalent to a unitary time and parity $PT$ transformation. All transformations of the proper orthochronous and antichronous transformations can be attributed to the meaning of actual transformations between two physical inertial frames \cite{Recami1982}.

\hspace{5mm} An antichronous Lorentz transformation $\Lambda_{+}^{\downarrow}$ changes the sign to the time-component of all four-vectors for a particle on the energy-momentum space, changing a positive energy particle moving forwards in time to a negative energy particle moving backwards in time. Thus, as shown in \cite{Debergh2018}, a proper $PT$ transformation implies negative masses on a time-like momentum four-vector $(mc, 0, 0, 0)$ for the rest frame, in which, for a particle of positive mass and positive time, a proper antichronous transformation changes to negative time and negative mass. Proper antichronous transformations, which result from the application of unitary $PT = (-diag(1, 1, 1, 1))$ transformations on orthochronous ones, join both positive time and mass, and negative time and mass regions of spacetime. 

\hspace{5mm} If relativistic observers can only explore their causally connected spacetime along the positive time axis, a reinterpretation of these negative energy particles moving backwards in time can be done so that they are observed as positive energy antiparticles travelling forward in time, as suggested in \cite{Recami1982}. But if their casually connected spacetime, where these negative energy particles exist, only contains negative energy particles and observers moving backwards in time, and no observation from positive time observers can be performed, this reinterpretation cannot take place.

\hspace{5mm} It is also reasoned in \cite{Souriau1997} in the context of the full Poincare group, that time inversions change the sign of a particle’s energy and mass: “\textit{time reversal changes the sign of the energy and thus, the sign of the mass. Consequently, it transforms every motion of a particle of mass m into a motion of a particle of mass $-m$.}”.

\hspace{5mm} One may also want to reject the proper antichronous transformations because they could allow causality-breaking paradoxes. These violations of causality occur only if both positive and negative masses are present in the same causally connected spacetime \cite{Bonnor1989}.

\hspace{5mm} The proposed negative time of a particle (equivalent to negative mass) is not a negative coordinate time $-t$, which would have a meaning relative to the Lorentz frame of an observer, but rather a negative proper time $-\tau$ as a property of the particle itself. A proper antichronous transformation $PT$ for a particle at rest leaves the term $c^2 d\tau^2 = ds^2$ invariant, but $d\tau=\sqrt{(dt^2)}=\pm dt$, and for every possible solution of $dt$, there exists an equally possible solution in which the negative sign for proper time is chosen. For the negative proper time case, in which its light cone points opposite (backwards) to that of positive proper time (forwards), it is a valid relativistic observer as it agrees with any other that the speed of light is the same in all inertial frames, and relative to other particles of negative proper time, all particles agree on their direction of time.

\hspace{5mm} Another reason that suggests that every unitary time transformation must be accompanied by a parity transformation, is that a negative Lorentz factor $\gamma = -1$ can be achieved by a negative proper time through $\gamma = dt/d \tau$, and also by a negative scaling along space dimensions by a negative length through $L_1 = L_0 / \gamma$, equivalent to the mirroring of the parity transformation. The same reasoning for negative Lorentz factors when applying $PT$ transformations has been followed in \cite{Villata2011}, but without considering negative masses. This negative value for the Lorentz factor could be obtained by generalizing its definition $\gamma^2 = 1/(1-\beta^2)$ so that $\gamma = \pm 1 / \sqrt{(1-\beta^2)}$.

\hspace{5mm} Consequently, proper antichronous transformations must be implemented into GR (which is derived from SR and thus, based on the restricted Lorentz group only) and considered physical in order to study the proposed time transformation at the EH and BH interior solution in a metric theory of gravity.

\hspace{5mm} In order to propose a time transformation at EHs, a global time coordinate must be fixed. This implies choosing a "preferred" frame of reference in which clocks show the "real" time. This can be done within the (experimentally indistinguishable and mathematically identical to SR) Lorentz aether theory interpretation of relativity, where time dilation is considered physical due to motion through the aether, covariance is broken because gravitational attraction changes to repulsion under a local change of time, and in which there exists a preferred reference frame: that in which the aether is at rest. A one-way speed of light experiment could in principle discern between both interpretations, although it is considered to be physically impossible to conduct. It is proposed that BHs serve the same purpose of this experiment, as they allow to set a preferred frame of reference in the proposed model.

\subsection*{2.5. Equivalence principles and the event horizon}

\hspace{5mm} The EP is often claimed to demand that any physical or noticeable change taking place at the EH of BHs is prohibited, because it would differentiate the effect of acceleration in flat space and the one experienced due to the gravitational field of the BH in a sufficiently small vicinity. A violation of the EP would imply an incompatibility with GR.

\hspace{5mm} For the proposed time transformation taking place at EHs of BHs which switches from gravity to antigravity, an infalling observer would only be able to detect the horizon through an internal gravitational experiment, so that Einstein’s equivalence principle (EEP): “\textit{The outcome of any local non-gravitational experiment in a freely falling laboratory is independent of the velocity of the laboratory and its location in spacetime}”, would hold. 

\hspace{5mm} What the proposed transformation would violate is the strong equivalence principle (SEP): “\textit{All test fundamental physics (including gravitational physics) is not affected, locally, by the presence of a gravitational field}”. The SEP can be thought of as an extension of EEP to gravitational phenomena \cite{DiCasola2015}, and satisfying the SEP is synonymous to consider that GR is the only possible theory of gravity. But EEP, by which fundamental non-gravitational test physics is not affected locally and at any point of spacetime by the presence of a gravitational field, is enough to formulate a metric theory of gravity in which fundamental non-gravitational physics in curved spacetime is locally Minkowskian. The SEP, by which the laws of gravitation are independent of velocity and location, would be violated in the modified theory of GR proposed to include proper antichronous transformations.

\subsection*{2.6. Time reversal symmetry in General Relativity}

\hspace{5mm} GR is symmetric under the classical definition of time-reversal which shifts momentum (the same definition of time reversal applied in the Feynman-Stüeckelberg interpretation). The equations of GR are based on the metric tensor with its quadratic form of the infinitesimal time coordinate $dt^2$, which is invariant under time reversal provided that the matter tensor is unchanged. Orbits around a massive body and particles attracted by it still orbit and are attracted after this time reversal. Even considering the case of a gravitationally accelerated particle in free fall to Earth which hits the ground in an inelastic collision, it is still attracted by the gravitational field when time reversed (the vibrational energy released in the collision drives the particle "upwards" when time is reversed, gravity does not accelerate up the particle but slows its upward velocity).

\hspace{5mm} In the standard Schwarzschild solution (the simplest BH, non-rotating and no electric charge), space flows with inward velocity and acceleration is also inward. It is invariant under the classical time reversal interpretation, since it is derived from a time-reversal symmetric differential equation and it is an equally valid solution when time-reversed: the BH transforms into a white hole (WH) in which mass is still positive so that acceleration is still inwards, but space flows with outwards velocity, and there is a horizon for which no particle can cross from outside to inside (an antihorizon).

\hspace{5mm} For both BH and WH, the horizons manifest a time asymmetry and thus, a preferred direction of time: future-timelike geodesics enter but they cannot emerge from them, and vice versa. For the process of a particle falling inside a BH EH, the reversed time process is a particle escaping the same horizon from inside. The contrary is prohibited by the same metric for both cases, and time must be reversed to switch from one to the other.

\hspace{5mm} In contrast, for the exterior negative mass standard Schwarzschild solution, the space flow velocity is imaginary and acceleration is outwards, which results in repulsive gravity and no horizon because there is no place where space flows at the at velocity equal to the speed of light. At the exterior solution, the ratio of proper time and Killing time $d\tau /dt$ increases while approaching the EH, so that time contraction instead of dilation takes place near the gravitational source, and the interior contains a naked singularity.

\hspace{5mm} Thus, even though a negative mass BH metric is obtained by a negative $r$ coordinate, which could be interpreted as a parity transformation (as it will be shown in the next section), and a WH metric is obtained by a negative $t$ coordinate, which could be interpreted as a time transformation, neither represent the interior solution proposed, since the first contains no horizons and the latter no repulsive gravitational interactions.

\subsection*{2.7. Metric solution}

\hspace{5mm} The proper time transformation explained in Sections 2.3 and 2.4 is consistent if applied only to the EH of BHs. This is possible since BH EHs are a globally and absolute defined boundary (to define and locate an EH one must know all the future history of spacetime), not defined with reference to any particular observer. Several indications in the current understanding of BHs in GR already point out that time transformations take place at EHs, which will be presented onwards.

\hspace{5mm} The suggested time transformation at the EH in the proposed model can be justified by the \textit{classical} and exact Schwarzschild metric solution formulated in 1916, built under spherical symmetry and time translation invariance, considering $t$ and $r$ time and space coordinates, and a constant determinant of the metric $det(g_{\mu v}) = -1$ with the sign convention $(+,-,-,-)$

\begin{equation}
	ds^2 = (1 - \frac{\alpha}{R}) dt^2 - \frac{dR^2}{1-\frac{\alpha}{R}} - R^2 d\theta^2 - R^2 (sin \theta)^2 d \varphi^2
\end{equation}

and the auxiliary quantity introduced by Schwarzschild $R$ following a non-linear relationship $R = \sqrt[3]{(r^3 + \alpha^3)}$ \cite{Schwarzschild1916}. For $r = 0, \ R = \alpha = 2GM$ there exists a singular EH, $M$ is just a constant of integration \cite{Bondi1957}, and $g_{tt} =0$. The solution, when applied to a simplified eternal BH model, does not cover the interior of the BH, since $r$ is not allowed to take negative values by definition and makes no assumptions about a specific interior geometry (or about any central singularities). David Hilbert’s modification to (3) considers $r = 0$ a coordinate singularity and substitutes $R \rightarrow r$, covering the interior and making the new $r = 0$ the central singularity, resulting in the \textit{standard} Schwarzschild solution. But Schwarzschild pointed out in \cite{Schwarzschild1916} that because $\frac{\alpha}{r} \approx 10^{-12}$ is very small for Mercury's orbit, one can use the Einstein's metric simplification $\frac{\alpha}{r}\rightarrow0$ and $R=r$ as an approximation for Mercury's perihelion precession estimation, although it is not the exact solution. Additionally, Einstein recognized that $g_{tt} = 0$ at the EH affirming that “\textit{This means that a clock kept at this place would go at the rate zero.}” \cite{Einstein1939}.

\hspace{5mm} As stated before, a physical time transformation cannot be applied to (3) because proper antichronous transformations as described in Section 2.4 are not included in GR. For having geodesic completeness in (3), one may allow $-r$ coordinates which leads to $-R$ in the metric if $-\alpha$ is considered for consistency in $R^3 = r^3 + \alpha^3$. This is equivalent to a change in the mass sign $M \rightarrow -M$, resulting in repulsive gravity in the negative mass BH exterior solution as explained in Section 2.2 \cite{Bonnor1989}. The $-r$ coordinate can be thought of as the negative solution of the conversion from Cartesian coordinates $r = \pm \sqrt{x^2 + y^2 + z^2}$, by which we are defining a new interior sphere for $ r< 0$, equivalent to parity  $P$ transformation, which does not correspond to any set of $(x, y, z)$ values from the Cartesian coordinates on the original $r> 0$ spacetime. The same reasoning has been followed for the hypothetical $-r$ interior of the Kerr metric in different coordinates in \cite{Villata2015}, which considers the transition to negative masses through the linear term $2Mr$ of the metric equivalent to a $PT$ transformation. A time transformation $T$ can accompany this parity transformation since it leaves the metric invariant. 

\hspace{5mm} A different interior can be glued to the exterior \textit{classical} Schwarzschild solution (3) under certain conditions such as the overall metric being continuous and differentiable at the joint and being the interior metric a valid solution to the field equations. An example of these extensions is described in \cite{Petit2015}, in which a shift of coordinates to the \textit{classical} Schwarzschild solution resulted in a $PT$ transformation at the EH, which the author argues is equivalent to a shift to attractive negative masses by an unjustified choice of negative mass interaction as explained in Section 2.2.

\hspace{5mm} Apart from this trivial consideration which relates a possible exterior solution extension of the \textit{classical} solution through a parity-time $PT$ transformation with negative masses, it is well known that the maximal extension to the \textit{standard} Schwarzschild metric through Kruskal–Szekeres coordinates describes two interior spacetime regions with singularities (II and IV quadrants), and two exterior spacetime regions with EHs (I and III quadrants). The quadrant I is the usual \textit{standard} Schwarzschild exterior solution in which a BH EH is present, and the quadrant III is an exterior solution where a WH EH is present. Related to these coordinates, it is argued in \cite{Bondarenko2019} that a mass inversion in the \textit{standard} Schwarzschild’s metric solution is counteracted by the discrete PT transformation of the coordinates in order to preserve the original dynamics. The discrete PT transformation of the coordinates and mass inversion in the \textit{standard} Schwarzschild’s metric solution is equivalent to the inversion of the Kruskal-Szekeres time and space coordinates. It is also shown in \cite{Grib2011} that within the exterior \textit{standard} Schwarzschild solution, a particle moving backwards in time is equivalent to have negative mass according to the timeline Killing vector and the conserved quantity $E^{\zeta}=mc^2(1-\frac{2GM}{rc^2})\frac{dt}{d\tau}$ which has the meaning of the energy of the particle.

\hspace{5mm} In the proposed model, both regions I and III are directly connected through a single horizon and region III is a closed spacetime. Consequently, both BH and WH horizons match sizes and the BH in-going energy is expelled directly as out-going energy by the WH, which grows in size in the process. When a positive mass crosses the EH, it is absorbed towards the interior as a negative mass.

\hspace{5mm} Due to the field equations and the symmetries that the Schwarzschild solution is built on, there are great constraints on what vacuum alternatives can be postulated inside so that the exterior is Schwarzschild. This is not problematic for the model proposed because all BHs are thought to have an stellar origin and thus, they always have a nonzero energy-momentum tensor since formation. More alternatives can be postulated if one considers time-dependent solutions, which are no longer static or Schwarzschild. 

\hspace{5mm} Consequently, there are substantial reasons to hypothesize that a unitary and proper antichronous transformation $PT$ takes place at EHs of BHs if proper antichronous transformations are allowed in a modified metric theory of gravity. This interior can now be characterized as a closed spacetime exclusively populated by negative energies and masses, in which a repulsive event antihorizon exists for which space flows with outwards velocity and acceleration is also outward, that can be interpreted as a negative mass WH. Light rays from the exterior crossing the horizon could be observed from the interior. A similar negative mass BH interior is described in \cite{Bonnor1989}, in which it is simplified as a fluid of negative energy density. Geodesics indicate that a massive test particle can never reach the central point of $r = 0$, solving the gravitational singularity with geodesic completeness through an energy condition violation of Penrose singularity theorems as explained in Section 2.1.

\hspace{5mm} The sign of the time coordinate would divide positive and negative energy and mass spacetimes, and the time dimension would have a defined ground state from which below it, time runs backwards. Near the horizon on either side, physical time flow slows down with respect to the ground state, but an observer could never measure this fundamental (and not relative) slowing of time in his local frame, because the same process of observation and data processing is synchronized with local spacetime, which is also slowed down in time. 

\hspace{5mm} One may think that if BH interiors consist of negative mass, they should repel other positive masses on their outside, such as other BHs or stars that orbit them (which contradicts observations). These negative mass BHs are described in \cite{Bonnor1989} and \cite{Mann1997}. But if the EH is responsible for switching the initial positive masses prior to the collapse of the BH into negative masses, it causally disconnects the interior from the exterior at the same time. Inside negative masses cannot influence the exterior geometry, which is remembered by the gravitational field with the corresponding positive curvature of positive mass from the original star prior to collapse, i.e., attractive gravity.

\hspace{5mm} Theoretical issues regarding the disparities between the interiors of standard Schwarzschild and Kerr BHs, such as those discussed in \cite{Villata2015}, would also be resolved: the rotating solution would lack singularities and inner Cauchy horizons, and it would differ from the static solution only in the sense that its interior content would be rotating.

\hspace{5mm} An interior solution which leaves the exterior solution unaltered might be considered meaningless to propose as it seems unfalsifiable. This is not the case, since any observer can travel inside the BH and perform an internal gravitational experiment to prove it. Nevertheless, a phenomenon might be explained by this model and a consequence of this interior by which it could be indirectly tested, will be discussed in the following section.

\section*{\bfseries 3. Discussion}\label{subsec:discussion}

\hspace{5mm} It could be hypothesized for a rotating BH that in the same way a time transformation takes places at the EH, a parity transformation should also take place at the ergosurface of the ergosphere due to spacetime flowing faster than the speed of light, so that both transformations must have always taken place inside the EH (since the EH always lies inside the ergosphere). Nothing suggests that this parity transformation could switch to negative masses (only the time transformation would), and therefore, for the EEP to hold, the parity transformation should not have any measurable physical effect on a test body (gravitational or non-gravitational). If a measurable physical effect could be attributed to that parity transformation (e.g., a charge transformation, a switch in sense of rotation, or a switch to mirror matter), the EEP (and the SEP) would be violated at the expense of being able to test the hypothesis and communicating infinitely far away the result of an experiment identifying the ergosurface. The unknown mechanisms of black hole astrophysical jets might be explained by one of these proposals inside ergospheres.

\hspace{5mm} The regular negative mass BH interior proposed suggests that the interior solution should undergo an antigravitational bounce instead of a collapse due to repulsive only gravity. Because the exterior is causally disconnected from the interior in one way (and assuming no accretion or BH merging), it can be then thought of as a homogeneous, isotropic and expanding cosmological spacetime with negative mass density.

\hspace{5mm} It could be argued that this inflation would increase the volume inside the BH without increasing its exterior surface area, leaving the exterior solution unaltered and leading to a case of Wheeler’s “bags of gold” solutions \cite{Wheeler1964} (eternal static BH exterior attached to an expanding FLRW interior). This solution to Einstein's equations contradicts the strong or "volume" interpretation of Bekenstein’s entropy equation $S=4\pi k_B G M^2 / c\hbar$ (and the holographic principle), by which entropy contained inside the volume is proportional to its surface area, and accounts for the number of total internal states of matter or micro-states, i.e., the information content of all the objects inside the hole is entirely encoded on its surface. For the strong form interpretation of Bekenstein’s entropy to hold in the proposed model (i.e., for preventing values of interior entropy larger than those allowed by the area law), the exterior surface area must also increase with the interior inflation process, in which volume (and entropy, since the number of possible scalar excitations grows with volume) increases, and the exterior EH radius would grow with time, which would be observed from outside as an increase of apparent mass. The rate of negative time inside the BH, which is slower near the inner antihorizon and faster away from it, would proportionally drive the increase in entropy, assuming that entropy does not decrease with negative time.

\hspace{5mm} Therefore, another mechanism of BH growth (independent from accretion and merging), is proposed to be simulated through an approximation of the time-dependent solution described in the previous sections and compared to observed and unexplained SMBHs sizes and apparent mass measurements up to $10^{10}$ solar masses. For small BHs, this inflationary period would be slowed down by interior gravitational time dilation for external faraway observers since masses inside are close to the inner antihorizon, explaining why newly formed BHs do not explode in size and we observe stellar mass BHs with masses close to the Tolman–Oppenheimer–Volkoff mass limit. But for massive stars which collapsed into BHs, such as population III stars from high redshift epochs, inflation would have occurred much faster, explaining the existence of observed SMBHs at high redshift, which hints that SMBHs grew quickly in the early age of the universe, an observation for which there is no accepted explanation. The first generation of stars, which are thought to have had between 100-1000 solar masses \cite{Ohkubo2009} when collapsed into BHs, would have grown faster than smaller ones, evolving into into SMBHs today. 

\hspace{5mm} The observed relationship between mass and spin, by which SMBHs rotate slower than smaller ones, suggests that their growth is not due to accretion (which drives BH spin), and could also be explained. The proposed model could also solve the unexplained intermediate mass gap between stellar BHs and SMBHs, being the intermediate mass black holes more sensitive to the described growth mechanism.

\hspace{5mm} Moreover, the proposed growth mechanism would slow down in time as inner mass density decreases, but it would never stop (ignoring a possible interior cosmological constant). If this never ending growth is enough to always counteract the BH decrease in size and mass due to Hawking radiation emission, which slows down as the EH surface grows, BH evaporation would be impossible, solving the information paradox without Hawking radiation having to carry any information away.

\hspace{5mm} Further study of the second law of thermodynamics at the interior, where time runs backwards, in relation to BH thermodynamics and entropy, would be interesting due to the well-known relationship of this law with the arrow of time for the possibility of decreasing entropy with negative time. This approach could also help address the negative energy problem in quantized gravity.

\section*{ \bfseries 4. Conclusion}

\hspace{5mm} A new singularity free black hole interior solution is briefly introduced which violates the energy conditions of the Penrose singularity theorems only at its interior, where exclusively gravitationally repulsive negative masses and negative energies are present. The considered antigravitational interaction between negative masses is based on the predictions of both Newtonian gravity and General Relativity, respecting Einstein's equivalence principle. It is argued that this is the only consistent possibility for the existence of negative masses, since the properties of black hole horizons naturally prevent their interaction with positive masses, avoiding the runaway motion, vacuum instability, and causality breaking paradoxes, which are the reason why negative masses have not been considered physical. 

It is shown that a shift to negative masses taking place at event horizons, does not violate Einstein's equivalence principle, and that time reversal is the most natural symmetry connecting positive and negative masses. This shift must be  the unitary parity-time PT transformation of relativistic quantum mechanics from the Feynman-Stüeckelberg interpretation of negative energy states in the Dirac equation, which is compatible with the proper antichronous transformation of the full Lorentz group. The time transformation must be physical and not a simple coordinate transformation, justified by a preferred frame of reference within the Lorentz aether interpretation of relativity, which is mathematically indistinguishable from Special Relativity. 

The metric solution is time-dependent and its interior undergoes an inflation due to repulsive gravity which is slowed down by time dilation, but increases the exterior black hole size due to increasing inner volume if the strong interpretation of Bekenstein's entropy holds. This results in another mechanism of black hole growth independent of accretion and merging, which could potentially explain supermassive black hole formation, the missing intermediate black hole mass gap, and solve the black hole information paradox by counteracting black hole evaporation due to Hawking radiation. A possible violation of the equivalence principle at ergosurfaces of rotating black holes due to a parity transformation is also speculated.

\def\bibsection{\section*{\refname}} 

\bibliography{references}

\end{document}